# Photonic radio frequency and microwave intensity differentiator based on an optical frequency comb source in an integrated micro-ring resonator

Xingyuan Xu,[1,a] Jiayang Wu,[1,a] Mehrdad Shoeiby,[2] Thach G. Nguyen,[2] Sai T. Chu,[3] Brent E. Little,[4] Roberto Morandotti,[5,6,7] Arnan Mitchell,[2] and David J. Moss[1,b]

[1]*Centre for Micro-Photonics, Swinburne University of Technology, Hawthorn, VIC 3122, Australia*

[2]*ARC Centre of Excellence for Ultrahigh-bandwidth Devices for Optical Systems (CUDOS), RMIT University, Melbourne, VIC 3001, Australia*

[3]*Department of Physics and Material Science, City University of Hong Kong, Tat Chee Avenue, Hong Kong, China.*

[4]*State Key Laboratory of Transient Optics and Photonics, Xi'an Institute of Optics and Precision Mechanics, Chinese Academy of Science, Xi'an, China.*

[5]*INSR-Énergie, Matériaux et Télécommunications, 1650 Boulevard Lionel-Boulet, Varennes, Québec, J3X 1S2, Canada.*

[6]*National Research University of Information Technologies, Mechanics and Optics, St. Petersburg, Russia.*

[7]*Institute of Fundamental and Frontier Sciences, University of Electronic Science and Technology of China, Chengdu 610054, China.*

We propose and experimentally demonstrate a microwave photonic intensity differentiator based on a Kerr optical comb generated by a compact integrated micro-ring resonator (MRR). The on-chip Kerr optical comb, containing a large number of comb lines, serves as a high-performance multi-wavelength source for implementing a transversal filter, which will greatly reduce the cost, size, and complexity of the system. Moreover, owing to the compactness of the integrated MRR, frequency spacings of up to 200-GHz can be achieved, enabling a potential operation bandwidth of over 100 GHz. By programming and shaping individual comb lines according to calculated tap weights, a reconfigurable intensity differentiator with variable differentiation orders can be realized. The operation principle is theoretically analyzed, and experimental demonstrations of first-, second-, and third-order differentiation functions based on this principle are presented. The radio frequency (RF) amplitude and phase responses of multi-order intensity differentiations are characterized, and system demonstrations of real-time differentiations for a Gaussian input signal are also performed. The experimental results show good agreement with theory, confirming the effectiveness of our approach.

## I. INTRODUCTION

With the ever-increasing demand for processing speed and throughput in modern communications systems, optical information processing technologies have attracted great interest due to their advantages in overcoming the intrinsic bandwidth bottleneck of electronic processing.[1–9] As one of the basic building blocks in optical signal processing and computing systems,[10] photonic differentiators are a key requirement in analyzing high-speed signals, as well as in waveform shaping, pulse generation, and systems control.[11-13]

---

[a]These authors contribute equally to this paper.

[b]Author to whom correspondence should be addressed. Electronic mail: dmoss@swin.edu.au

To implement photonic differentiators, a number of schemes have been proposed, which can be classified into two categories - namely, field differentiators and intensity differentiators.[14] Field differentiators based on apodized fibre Bragg gratings[11,13,15] and integrated silicon photonic devices[16-20] have recently been demonstrated. These types of devices yield the derivative of a complex optical field, and have the ability to shape ultra-short optical pulses that could find applications in optical pulse generation and advanced coding.[21–23] On the other hand, some other applications such as ultra-wideband frequency generation, radio frequency (RF) measurement and filters, require intensity differentiators that provide the derivative of the temporal intensity profiles associated with RF signals.[24–26] A photonic intensity differentiator based on a dual-drive Mach-Zehnder modulator together with an RF delay line, was reported [14] but the processing speed was intrinsically limited by the operation bandwidth of the RF delay line. Photonic intensity differentiators based on semiconductor optical amplifiers (SOAs) and optical filters (OFs) have also been reported,[27,28] featuring high processing speeds of up to 40-Gb/s. This approach, however, works only for a fixed differentiation order and lacks reconfigurability, whereas in practical applications processing systems with variable differentiation orders are desired to meet diverse computing requirements. To implement highly reconfigurable intensity differentiators, transversal filter schemes based on discrete microwave photonic delay-lines have been investigated.[12, 29] However, these approaches have had limitations of one form or another, such as the need for generating the taps using discrete laser arrays, thus significantly increasing the system cost and complexity.

In this paper, a reconfigurable microwave photonic intensity differentiator based on an integrated Kerr optical frequency comb source is proposed and experimentally demonstrated. By employing an on-chip nonlinear micro-ring resonator (MRR), we generate a broadband Kerr comb with a large number of lines, and use it as a high-quality multi-wavelength source for implementing a transversal filter. Moreover, the large frequency spacing of the integrated Kerr comb source also yields an increased Nyquist zone, thus leading to a potential operational bandwidth of over 100 GHz - well beyond the processing bandwidth of electronic devices. By programming and shaping the power of individual comb lines according to a set of corresponding tap weights,[30, 31] reconfigurable intensity differentiators with variable differentiation orders can be achieved. We present a detailed analysis of the operation principle, and perform experimental demonstrations of first-, second-, and third-order intensity differentiators using the fabricated device. The RF amplitude and phase response of the intensity differentiator are experimentally characterized, and systems demonstrations of real-time differentiators for Gaussian pulse input signals are also carried out. The experimental results are consistent with theory, corroborating the feasibility of our approach as a solution to implement high-speed reconfigurable microwave photonic intensity differentiators.

## II. OPERATION PRINCIPLE

Based on the classical theory of signals and systems,[32] the spectral transfer function of an $N^{th}$ order temporal differentiator can be expressed as

$$H(\omega) \propto (j\omega)^N, \qquad (1)$$

where $j = \sqrt{-1}$, $\omega$ is the angular frequency, and $N$ is the differentiation order. According to the above transfer function, the amplitude response of a temporal differentiator is proportional to $|\omega|^N$, while the phase response has a linear profile, with a zero and $\pi$ jump at zero frequency for $N$ even and odd, respectively. The ideal RF amplitude and phase response of first-, second-, and third-order microwave differentiators are shown in Figs. 1(a)–(c), respectively.

In this paper, we employ a versatile approach towards the implementation of microwave photonic differentiators based on transversal filters, where a finite set of delayed and weighted replicas of the input RF signal are produced in the optical domain and combined upon detection.[33–35] The transfer function of a typical transversal filter can be described as

$$H(\omega) = \sum_{n=0}^{M-1} a_n e^{-j\omega nT}, \qquad (2)$$

where $M$ is the number of taps, $a_n$ is the tap coefficient of the $n$-th tap, and $T$ is the time delay between adjacent taps. It should be noted that differentiators based on Eq. (2) are an intensity differentiators for baseband RF input signals, i.e., the combined output RF signal after detection yields an exact differentiation of the input RF signal, in contrast to field differentiators that yield the derivative of a complex optical field.[11,13,15-20] We note that, while optical field differentiators can be used to directly operate on microwave photonic signals, our approach has important advantages. When an RF signal is modulated onto an optical carrier, the intensity of the optical carrier is proportional to the square of the RF field. Thus, techniques that differentiate the optical field will yield the derivative of the square of the RF function rather than the exact derivative of the RF function directly, as our technique does.

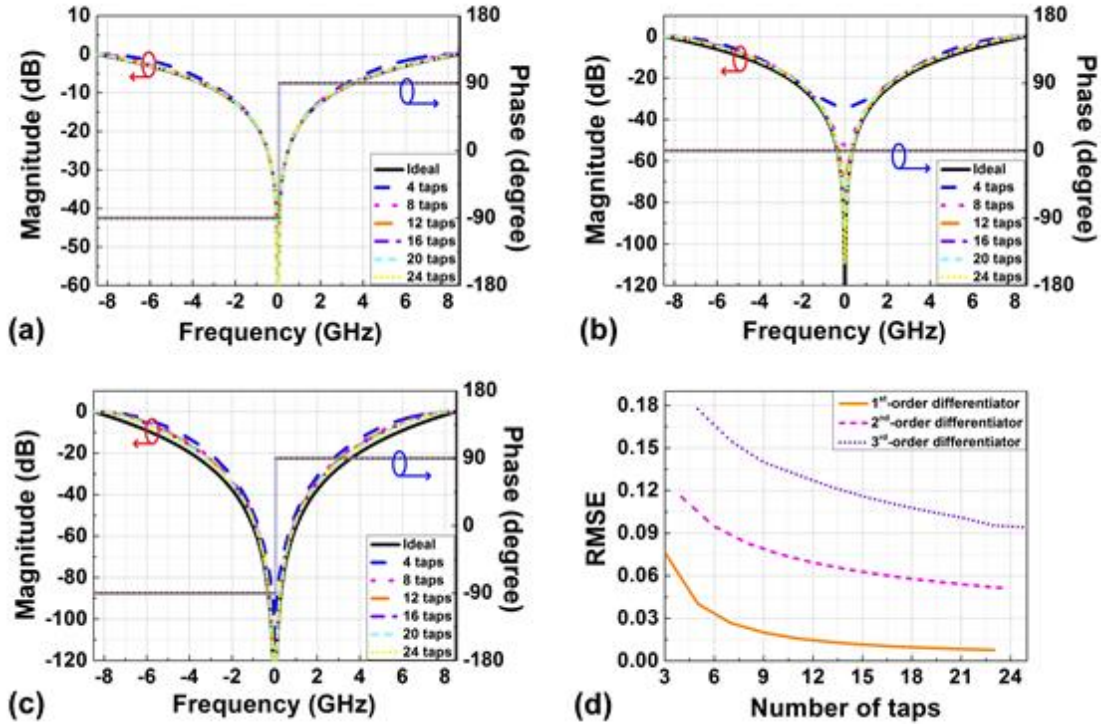

Fig. 1. Simulated RF amplitude and phase response of the (a) first-, (b) second-, and (c) third-order temporal differentiators. The amplitude and phase response of the differentiators designed based on Eq. (2) are also shown according to the number of taps employed. (d) RMSEs between the calculated and ideal RF amplitude response of the first-, second-, and third-order intensity differentiators as a function of the number of taps.

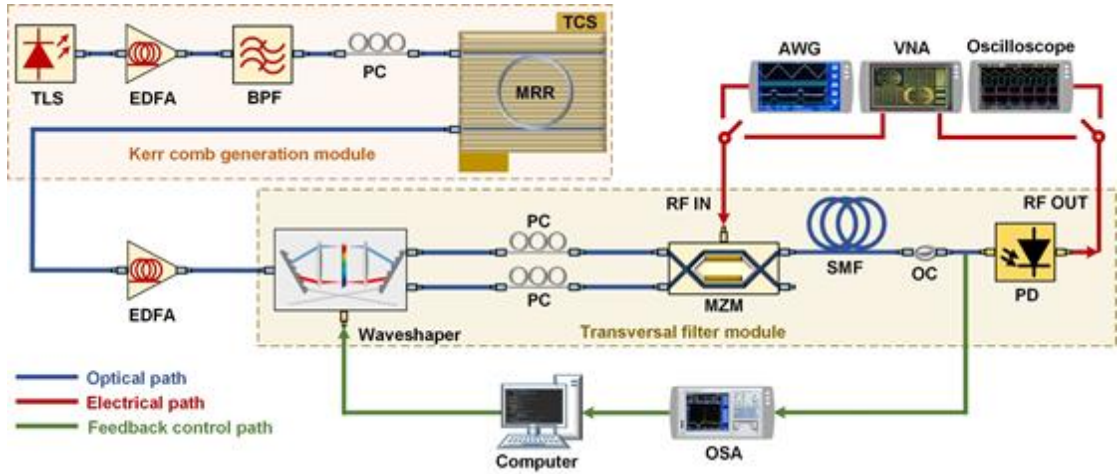

Fig. 2. Schematic illustration of the proposed reconfigurable microwave photonic intensity differentiator. TLS: tunable laser source. EDFA: erbium-doped fibre amplifier. PC: polarization controller. BPF: optical bandpass filter. TCS: temperature control stage. MZM: Mach-Zehnder modulator. SMF: single mode fibre. OC: optical coupler. PD: photo-detector. OSA: optical spectrum analyzer. VNA: vector network analyzer. AWG: arbitrary waveform generator.

To implement the temporal differentiator in Eq. (1), we calculate the tap coefficients in Eq. (2) based on the Remez algorithm.[36] The corresponding amplitude and phase response of the first-, second-, and third-order differentiators as a function of the numbers of taps are also plotted in Figs. 1(a)–(c). When the number of taps is increased, it is clear that the discrepancies between the amplitude response of the transversal filters and the ideal differentiators are improved for all three orders, whereas the phase response of the transversal filters is identical to that of the ideal differentiators regardless of the number of taps. To quantitatively analyze the discrepancies in the amplitude responses, we further calculate the root mean square errors (RMSEs) for the first-, second-, and third-order differentiators as a function of the number of taps (see Fig. 1(d)). One can see that the RMSE is inversely proportional to the number of taps, as reasonably expected. In particular, we note that when the number of taps increases, the RMSE decreases dramatically for a small number of taps, and then decreases more gradually as the number of taps becomes larger.

Figure 2 shows a schematic illustration of the reconfigurable microwave photonic intensity differentiator. It consists of two main blocks: the Kerr optical frequency comb generation module based on a nonlinear MRR and a transversal filter module for reconfigurable intensity differentiation. In the first module, the continuous-wave (CW) light from a tunable laser source is amplified by an erbium-doped fibre amplifier (EDFA), followed by a tunable optical bandpass filter (BPF) to suppress the amplified spontaneous emission (ASE) noise. A polarization controller (PC) is inserted before the nonlinear MRR to make sure that the polarization state matches the desired coupled mode. When the wavelength of the CW light is tuned to a resonance of the nonlinear MRR and the pump power is high enough for sufficient parametric gain, the optical parametric oscillation (OPO) process in the nonlinear MRR is initiated, generating a Kerr optical comb with nearly equal line spacing.[37,38] The nonlinear MRR is mounted on a highly precise temperature control stage (TCS) to avoid thermal resonance drift and to maintain the wavelength alignment of the resonance to the CW pump light. Owing to the compact size and ultra-high quality factor of the nonlinear MRR, the generated Kerr comb provides a large number of wavelength channels with narrow linewidths for the subsequent transversal filter module. With respect to conventional intensity differentiators based on laser diode arrays, the cost, size and complexity can be greatly reduced. After being amplified by another EDFA, the generated Kerr comb is directed to the second module where it is processed by a waveshaper to get weighted taps according to the coefficients calculated by means of the Remez algorithm. Considering that the generated Kerr comb is not flat, a real-time feedback control path is introduced to read and shape the comb lines' power accurately. A 2×2 balanced Mach-Zehnder modulator (MZM) is employed to generate replicas of the input RF signal. When the MZM is quadrature-biased, it can simultaneously modulate the input RF signal on both positive and negative slopes, thus yielding modulated signals with opposite phases as well as tap coefficients with opposite algebraic signs. After being modulated, the tapped signals from one output of the MZM are delayed by a dispersive fibre. The time delay between adjacent taps is determined jointly by the frequency spacing of the employed comb source and the dispersion accumulated in the fibre. Finally, the weighted and delayed taps are combined upon detection and converted back into RF signals to form the differentiation output.

It is worth mentioning that due to the intrinsic advantages of transversal filters, our scheme features a high degree of reconfigurability in terms of processing functions and operation bandwidth, thus offering a reconfigurable platform for diverse microwave photonic computing functions. By simply programming the waveshaper to shape the comb lines according to the corresponding tap coefficients, our scheme can also apply to other computing functions such as Hilbert transforms and differential equation solving.[39,40] Note that the high reconfigurability of the proposed differentiator cannot typically be achieved by passive silicon counterparts[16-20], thus making our approach more suitable for diverse computing requirements in practical applications. The operation bandwidth can also be changed by adjusting the time delay between adjacent taps or employing different tap coefficients. An increased operation bandwidth can be obtained by simply employing a dispersive fibre with a shorter length. The operation bandwidth is fundamentally limited by the Nyquist zone, which is determined by the comb spacing. In our case, the frequency spacing of the Kerr comb generated by the nonlinear MRR reaches 200 GHz, thus leading to a potential operation bandwidth of over 100 GHz, which is well beyond electrical processing bandwidths and comparable with that associated with integrated-waveguide Bragg gratings.[20]

## III. EXPERIMENTAL RESULTS

In our experiment (see Fig. 3(a)), the nonlinear MRR used to generate the Kerr comb was fabricated on a high-index doped silica glass platform using CMOS compatible fabrication processes.[37,38,41-43] First, high-index (n = ~1.70 at 1550 nm) doped silica glass films were deposited using standard plasma enhanced chemical vapour deposition (PECVD), then photolithography and reactive ion etching (RIE) were employed to form waveguides with exceptionally low surface roughness. Finally, silica glass (n = ~1.44 at 1550 nm) was deposited via PECVD as an upper cladding. Our CMOS compatible fabrication process makes our differentiators comparable to, in terms of fabrication maturity, those implemented by means of optoelectronic silicon devices[16-20]. In particular we note that, due to the ultra-low loss of our platform, the ring resonator has a quality factor of ~1.2 million. Our device architecture uses a vertical coupling scheme where the gap can be controlled via film growth - a more accurate approach than lithographic techniques[40,44]. The gap between the bus waveguide and the MRR is approximately 200nm. The compact integrated MRR has a radius of ~135 µm with a relatively large free spectral range (FSR) of ~1.6 nm, i.e., ~200 GHz. Such a large FSR enables an increased Nyquist zone of ~100 GHz, which is challenging for mode-locked lasers and externally-modulated comb sources.[46–48] The advantages of our platform for nonlinear OPOs include ultra-low linear loss (~0.06 dB·$cm^{-1}$), a moderate nonlinearity parameter (~233 $W^{-1}·km^{-1}$), and in particular a negligible nonlinear loss up to extremely high intensities (~25 GW·$cm^{-2}$).[37,38,41-43] After packaging the input and output ports of the device with fibre pigtails, the total insertion loss is ~3.5 dB. A scanning electron microscope (SEM) image of the cross-section of the MRR before depositing the $SiO_2$ upper cladding is shown in Fig. 3(b). By boosting the power of the CW light from the tunable laser source via an EDFA and adjusting the polarization state, multiple FSR mode-spaced combs were first generated, in which the primary spacing was determined by the parametric gain. When the parametric gain lobes became broad enough, secondary comb lines with a spacing equal to the FSR of the MRR were generated via either degenerate or non-degenerate four wave mixing (FWM). In our experiment, the power threshold for the generation of secondary comb lines was ~500 mW. The resulting Type II Kerr optical comb[49] ( Fig. 4(a) ) was over 200-nm wide, and flat over ~32 nm. Since the generated comb only served as a multi-wavelength source for the subsequent transversal filter, in which the optical power from different taps was detected incoherently by the photo-detector, achieving rigorous comb coherence was not crucial and the proposed differentiator was able to work under relatively incoherent conditions. In the experiment, the numbers of taps used for first-, second-, and third-order differentiation demonstrations were 8, 6, and 6, respectively. The choice of these numbers was made mainly by considering the power dynamic range, i.e., the difference between the maximum power of the generated comb lines and the power associated with the noise floor. The dynamic range was determined by the EDFA before waveshaping, which in our case was ~30 dB. An increased number of taps requires a broader power dynamic range, which can be achieved by using an EDFA with a lower noise floor. As analysed in section Ⅱ, more taps are needed when the differentiation order increases, and for a fixed number of taps, increasing the order of differentiation also increases the required power dynamic range. In order to get better performance with a limited number of taps, we decreased the operation bandwidth of the second- and third-order differentiators to half that of the transversal filter's Nyquist frequency when engineering the response function with the Remez algorithm. It should be noted that the actual bandwidth of the differentiator is not limited by this design since the FSR of the transversal filter can be increased. The calculated tap coefficients for first-, second-, and third-order differentiations are listed in Table I. The selected comb lines of the generated optical comb were processed by the waveshaper based on these coefficients. Considering that the generated Kerr comb was not flat or absolutely stable, we adopted a real-time feedback control path to increase the accuracy of comb shaping. The comb lines' power was first detected by an optical spectrum analyzer (OSA) and compared with the ideal tap weights. Subsequently, an error signal was generated and fed back into the waveshaper to calibrate the system and achieve accurate comb processing. The shaped optical combs are shown in Figs. 4(b)–(d). A good match between the measured comb lines' power (red solid line) and the calculated ideal tap weights (green crossing) was obtained, indicating that the comb lines were successfully shaped. They were then divided into two parts according to the algebraic sign of the tap coefficients and fed into the 2×2 balanced MZM biased at quadrature. The modulated signal after the MZM was propagated through ~2.122-km single mode (dispersive) fibre (SMF). The dispersion of the SMF was ~17.4 ps/(nm· km), which corresponds to a time delay of ~59 ps between adjacent taps and yielded an effective FSR of ~16.9 GHz in the RF response spectra.

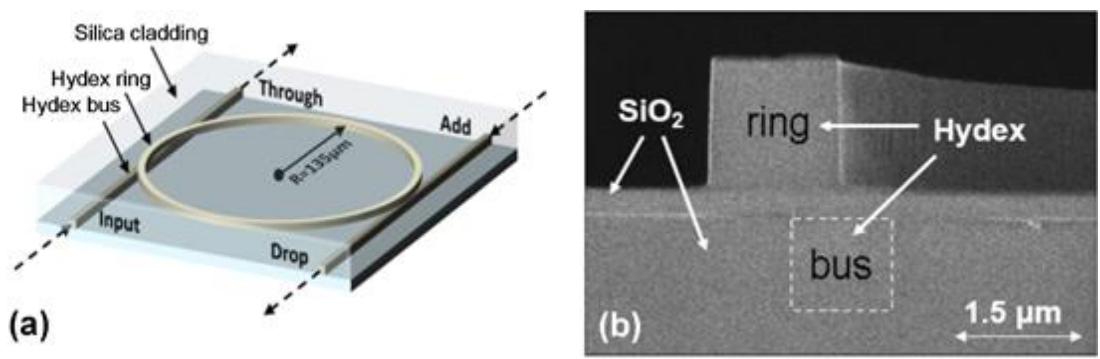

Fig. 3. (a) Schematic of the micro-ring resonator. (b) SEM image of the cross-section of the resonator before depositing the upper cladding.

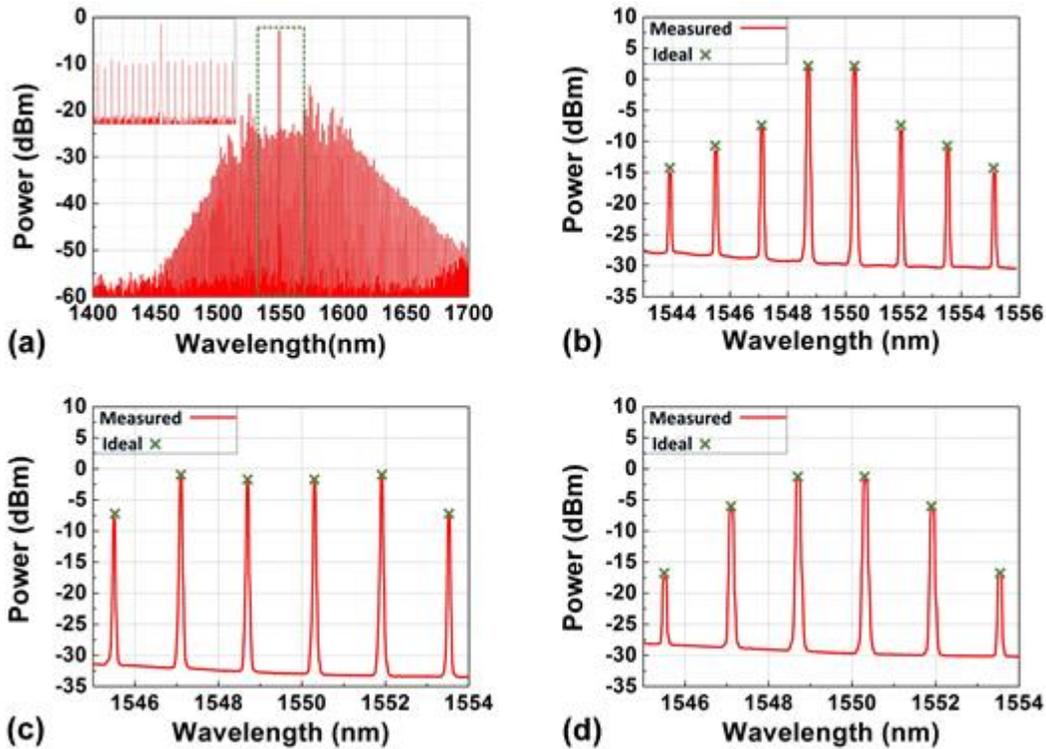

Fig. 4. (a) Optical spectrum of the generated Kerr comb in a 300-nm wavelength range. The inset shows a zoom-in spectrum with a span of ~32 nm. (b)─(d) Measured optical spectra (red solid) of the shaped optical combs and ideal tap weights (green crosses) for the first-, second-, and third-order intensity differentiators.

**TABLE I. Tap coefficients for the first-, second-, and third-order differentiations**

| Order of differentiation | Number of taps | Tap coefficients |
|---|---|---|
| First-order | 8 | [−0.0226, 0.0523, −0.1152, 1, −1, 0.1152, −0.052, 0.0226] |
| Second-order | 6 | [0.0241, −0.1107, 0.0881, 0.0881, −0.1107, 0.0241] |
| Third-order | 6 | [0.0450, −0.4076, 1, −1, 0.4076, −0.0450] |

After the weighted and delayed taps were combined upon detection, the RF response for different differentiation orders was characterized by a vector network analyser (VNA, Anritsu 37369A). Figures. 5(a-i), (b-i), and (c-i) show the measured and simulated amplitude response of the first-, second-, and third-order intensity differentiators, respectively. The corresponding phase responses are depicted in Figs. 5(a-ii), (b-ii), and (c-ii). It can be seen that all three configurations exhibit a response expected from ideal differentiation. In Fig. 5(a-i), we also indicate the operating frequency range of the first-order intensity differentiator. Since our device was designed to perform intensity differentiation for baseband RF signals, the operating frequency range starts at DC and ends at half of the spectral range between DC and the notch centred at ~17 GHz. The FSR of the RF response spectra is ~16.9 GHz, which is consistent with the time delay between adjacent taps. Note that by adjusting the FSR of transversal filter through the dispersive fibre or by programming the tap coefficients, a variable operation bandwidth for the intensity differentiator can be achieved, which is advantageous for meeting diverse requirements.

We also performed system demonstrations of real-time signal differentiation for baseband Gaussian-like input pulses with a full-width at half maximum (FWHM) of ~0.12 ns, generated by an arbitrary waveform generator (AWG, KEYSIGHT M9505A), as shown in Fig. 6(a). The waveform of the output signals after differentiation are shown in Figs. 6(b)–(d) (blue solid curves). They were recorded by means of a high-speed real-time oscilloscope (KEYSIGHT DSOZ504A Infinium). For comparison, we also depict the ideal differentiation results, as shown in Figs. 6(b)–(d) (red dashed curves). The experimental Gaussian pulse in Fig. 6(a) is used as the input RF signal for the simulation. One can see that the measured curves closely match their theoretical counterparts, indicating good agreement between experimental results and theory. Unlike the field differentiators [11, 13, 15–20], the temporal derivatives of intensity profiles can be observed, indicating that intensity differentiation was successfully achieved. For the first-, second-, and third-order differentiators, the calculated RMSE between the measured and the theoretical curves are ~4.15%, ~6.38%, and ~7.24%, respectively.

To further investigate the imperfections associated with the device performance, we employed commercial software (VPI photonics) to simulate the RF amplitude and phase response of the various differentiators by considering the tap weight error during the comb shaping, the chirp induced by the MZM, and the third-order dispersion (TOD) of the fibre. Based on the measured error signal from the feedback control path and the empirical chirp/TOD values in previous experiments[39], the tap weight error, chirp coefficient, and TOD in our simulation were set to 0.5 dB, 0.5, and 0.083 ps/(nm$^2$· km), respectively. The simulated amplitude and phase response are plotted in Figs. 5(a)–(c). One can see that after incorporating these effects, the simulated RF responses fit more closely to the experimental results, thus confirming that the degradation of the RF response can be attributed to all these effects. Note that although there are phase jumps in the notches centred at DC in Figs. 5(a)–(c), they only occur in relatively narrow bandwidths, and therefore do not significantly affect the differentiation performance. Similar phase jumps were also observed in our previous experiments.[39] By employing a low-chirped MZM, further increasing the resolution of the feedback control path, and compensating the TOD, we can anticipate improved performance of the differentiators.

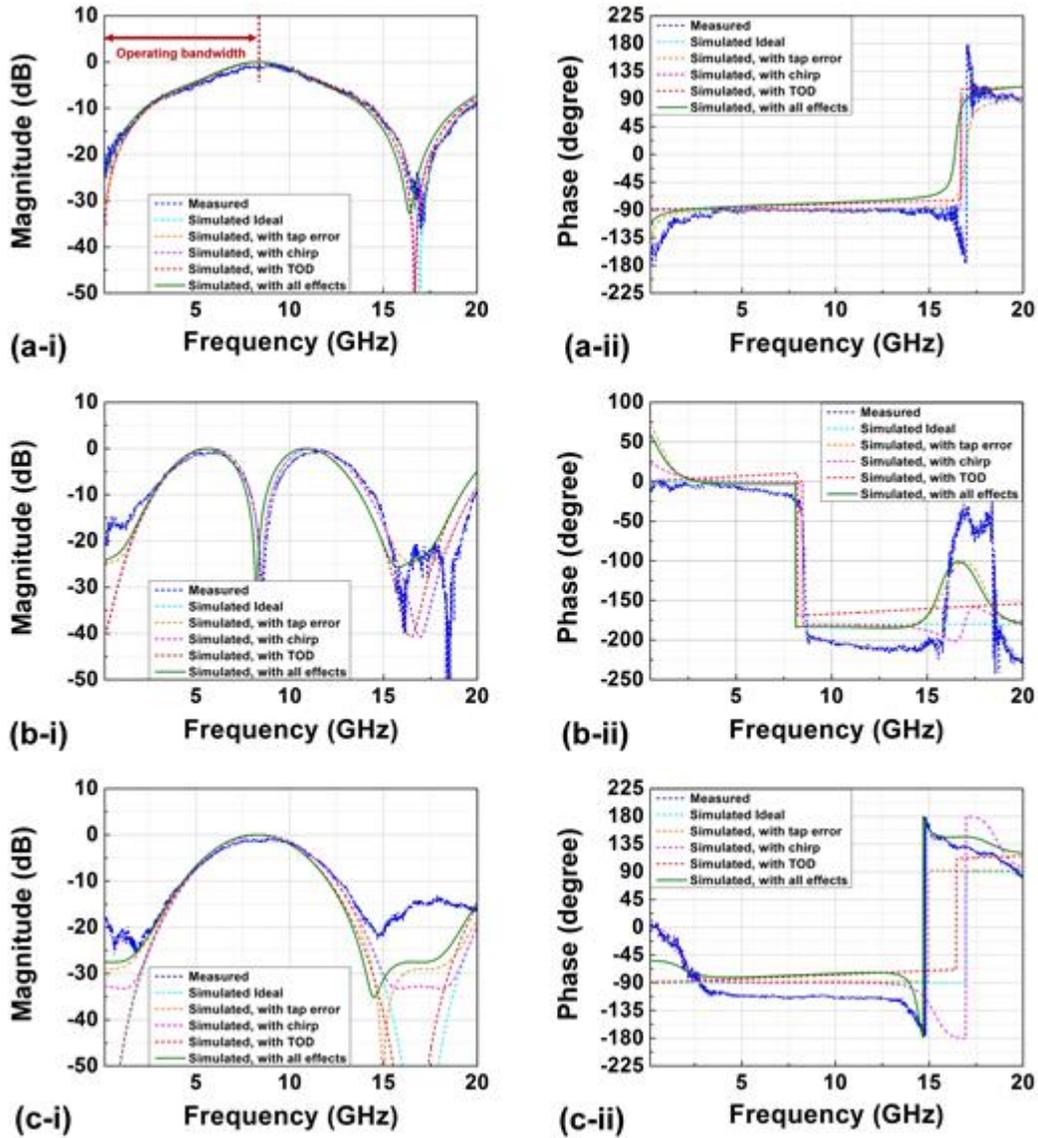

Fig. 5. Measured and calculated RF amplitude and phase response of (a-i)–(a-ii) first-order, (b-i)–(b-ii) second-order, and (c-i)–(c-ii) third-order intensity differentiators. The simulated amplitude and phase response after incorporating the tap error, chirp, and the third-order dispersion (TOD) are also shown accordingly. The operating frequency range is also indicated.

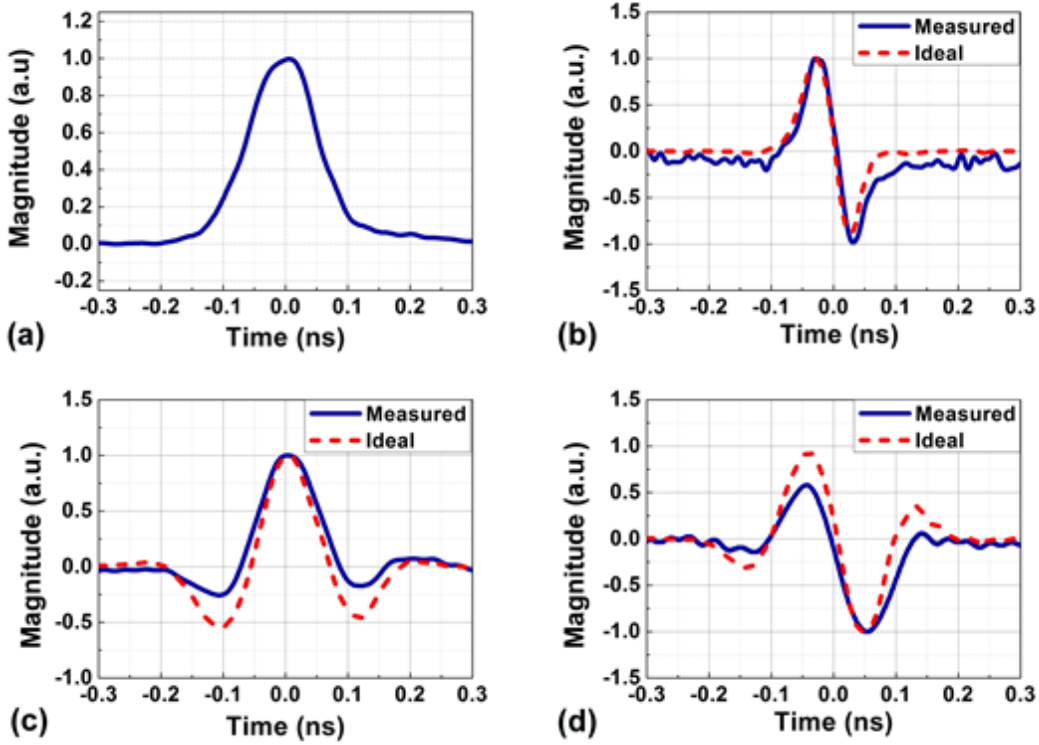

Fig. 6. (a) Measured temporal waveform of a Gaussian input pulse. Theoretical (red dashed) and experimental (blue solid) responses of the (b) first-, (b) second-, and (c) third-order intensity differentiators.

## IV. CONCLUSION

We propose and demonstrate a reconfigurable microwave photonic intensity differentiator based on an integrated Kerr comb source. By programming and shaping the individual comb line powers according to calculated tap weights, we successfully demonstrate first-, second-, and third-order intensity differentiation of RF signals. The RF amplitude and phase responses of the proposed differentiator are characterized, and systems demonstrations of real-time differentiations are performed for Gaussian input pulses. We achieve good agreement between theory and experiment, thus verifying the effectiveness of our approach. Our technique, based on a CMOS-compatible nonlinear micro-ring resonator, provides a new way to implement microwave photonic intensity differentiators featuring compact device footprint, high processing bandwidth, and high reconfigurability, thus holding great promise for future ultra-high-speed computing and information processing.

**ACKNOWLEDGMENTS**

This work was supported by the Australian Research Council Discovery Projects Program (No. DP150104327). RM acknowledges support by Natural Sciences and Engineering Research Council of Canada (NSERC) through the Strategic, Discovery and Acceleration Grants Schemes, by the MESI PSR-SIIRI Initiative in Quebec, and by the Canada Research Chair Program. He also acknowledges additional support by the Government of the Russian Federation through the ITMO Fellowship and Professorship Program (grant 074-U 01) and by the 1000 Talents Sichuan Program in China. Brent E. Little was supported by the Strategic Priority Research Program of the Chinese Academy of Sciences, Grant No. XDB24030000.